\documentclass[twocolumn]{aastex62}

\usepackage{natbib}
\usepackage{amsmath}
\usepackage{multirow}
\usepackage[normalem]{ulem}
\usepackage{xcolor}

\definecolor{tropicalrainforest}{rgb}{0.0, 0.46, 0.37}

\shorttitle{Giant Planet Turnover}
\shortauthors{Fernandes et al.}

\begin{document}

\title{Hints for a Turnover at the Snow Line in the Giant Planet Occurrence Rate}

\correspondingauthor{Rachel B. Fernandes}
\email{rachelbf@lpl.arizona.edu}

\author[0000-0002-3853-7327]{Rachel B. Fernandes}
\affil{Lunar and Planetary Laboratory, The University of Arizona, Tucson, AZ 85721, USA}
\affil{Earths in Other Solar Systems Team, NASA Nexus for Exoplanet System Science}

\author[0000-0002-1078-9493]{Gijs D. Mulders}
\affil{Department of the Geophysical Sciences, The University of Chicago, Chicago, IL 60637, USA}
\affil{Earths in Other Solar Systems Team, NASA Nexus for Exoplanet System Science}

\author[0000-0001-7962-1683]{Ilaria Pascucci}
\affil{Lunar and Planetary Laboratory, The University of Arizona, Tucson, AZ 85721, USA}
\affil{Earths in Other Solar Systems Team, NASA Nexus for Exoplanet System Science}

\author[0000-0002-1013-2811]{Christoph Mordasini}
\affil{Physikalisches Institut, Universit\"{a}t Bern, Gesellschaftsstrasse 6, CH-3012 Bern, Switzerland}

\author[0000-0002-8811-1914]{Alexandre Emsenhuber}
\affil{Lunar and Planetary Laboratory, The University of Arizona, Tucson, AZ 85721, USA}
\affil{Physikalisches Institut, Universit\"{a}t Bern, Gesellschaftsstrasse 6, CH-3012 Bern, Switzerland}

\begin{abstract}
The orbital distribution of giant planets is crucial for understanding how terrestrial planets form and predicting yields of exoplanet surveys. Here, we derive giant planets occurrence rates as a function of orbital period by taking into account the detection efficiency of the \textit{Kepler} and radial velocity (RV) surveys. The giant planet occurrence rates for \textit{Kepler} and RV show the same rising trend with increasing distance from the star. We identify a break in the RV giant planet distribution between $\sim$2-3\,au --- close to the location of the snow line in the Solar System --- after which the occurrence rate decreases with distance from the star. Extrapolating a broken power-law distribution to larger semi-major axes, we find good agreement with the $\sim 1\%$ planet occurrence rates from direct imaging surveys. Assuming a symmetric power law, we also estimate that the occurrence of giant planets between 0.1$-$100\,au is $26.6^{+7.5}_{-5.4}\%$ for planets with masses 0.1-20\,$M_{\rm J}$ and decreases to $6.2^{+1.5}_{-1.2}\%$ for planets more massive than Jupiter. This implies that only a fraction of the structures detected in disks around young stars can be attributed to giant planets. Various planet population synthesis models show good agreement with the observed distribution, and we show how a quantitative comparison between model and data can be used to constrain planet formation and migration mechanisms.
\end{abstract}


\section{Introduction} \label{sec:intro}
Giant planets, hereafter GPs, form while substantial gas is still present in disks, i.e. within ~10 Myr (e.g., \citealt{pascucci2006}). In the standard picture of planet formation (e.g., \citealt{raymond2005}), terrestrial planets take much longer to form, of the order of hundreds of millions of years. As such, the presence, mass, and eccentricity of GPs directly impact the final location and mass of terrestrial planets (e.g., \citealt{levison2003}). For example, \citet{raymond2006} finds that GPs inside roughly 2.5\,au inhibit the growth of 0.3\,M$_\earth$ planets in the habitable zone of sun-like stars. In addition, GPs affect the delivery of water to terrestrial planets (e.g., \citealt{morbidelli2012}), a key ingredient for the development of life as we know it.

The observed distribution of giant planets provides important clues to how and when they form. GPs preferentially form beyond the snow line since their formation is expected to be more efficient there due to an increased amount of solids as water vapor condenses onto ice (e.g., \citealt{kennedy2008}). In disks around young solar analogues, the snow line is expected to be between $\sim$2-5\,au (e.g., \citealt{mulders2015snow}). For the Solar System, it is inferred to be at $\sim$2.5\,au at the time of planetesimal formation from the gradient in composition of large main belt asteroids (e.g., \citealt{demeo2014}). If indeed GPs preferentially form beyond the snowline, those detected within $\sim$1\,au underwent significant migration, either through interaction with the gas disk (e.g., \citealt{goldreich1980,lin1986,lin1996}) or as a result of planet scattering (e.g., \citealt{rasio1996,weidenschilling1996,ford2006,chatterjee2008}).

Alternatively, some theorists have argued that a substantial fraction of the hot and warm Jupiters could have formed in situ  (e.g. see \citealt{batygin2016,boley2016,bailey2018}).

Core-accretion planet formation models that include disk migration typically predict that the occurrence of giant planets increases with distance from the star within $\sim$1\,au (e.g.,  \citealt{ida2004b,mordasini2009}) but often show a break at larger distances. 
The exact shape of the distribution depends on the physics of planet formation: how planet cores grow and interact (e.g., \citealt{mordasini2018}), how quickly planets migrate through the disk (e.g., \citealt{ida2008_5,ida2018}), and the timescale and mechanism by which the disk disperses (e.g., \citealt{alexander2012deserts}). While superficial comparisons between these models and the detected GP population have been made (e.g., see \citealt{ida2004a,ida2004b}), a detailed statistical analysis in which survey completeness was taken into account has not been done.

The \textit{Kepler} mission has provided detailed exoplanet population statistics for a large range of planet sizes close to their host stars (e.g., \citealt{howard2012}). 
GPs show a rising occurrence rate out to the $\sim$1\,au semi-major axis covered by {\it Kepler} \citep{dongzhu13,santerne2016}.
The radial velocity (hereafter RV) technique extends exoplanet detections well beyond $\sim 1$\,au, but only for planets more massive than Neptune (e.g., \citealt{howard2010,mayor2011,wittenmyer2016}). 

The analysis of these RV data has established that GPs are much rarer than the Neptune and Super-Earths detected by {\it Kepler} with an occurrence of only $\sim$10\% within a few years (see e.g., Table 1 in \citealt{winn2015}). Early studies have also shown that the RV occurrence rate could be described by a power law in planet mass and orbital period for GPs 0.3-10\,$M_{\rm J}$ inside 2,000 days \citep[e.g.,][]{cumming2008}. The frequency of these GPs was found to decrease with increasing mass and increase with period. However, direct imaging surveys recognized early on that such power law could not extend to the large orbital separations this technique is sensitive to, beyond $\sim$30\,au, as they detected very few, if any, exoplanets (e.g., \citealt{kasper2007}).
Larger imaging surveys with improved instrumentation and analysis techniques are confirming that the frequency of GPs on wide orbits is indeed low, $\leq$\,1\% (e.g., \citealt{bowler2016,galicher2016}).
Additionally, RV trend studies have been important to  bridge the gap between the population of close-in planets detected via RV and the further way one discovered by direct imaging \citep{montet2014,knutson2014,bryan2016,bryan2018}.
Particularly relevant to our study is \citet{bryan2016} who suggest a declining frequency of giants already beyond 3-10\,au.

Here, we first compare the \textit{Kepler} GP occurrence rate from the latest data release with the RV occurrence from \citet{mayor2011} corrected for the survey completeness (Section \ref{sub:rv}). We show that beyond 10\,days the occurrence of planets with masses 0.1-20\,M$_{\rm J}$ (radii $> 5$\,R$_\earth$) match well, meaning that the RV GP occurrence rate  can be used to extend that from \textit{Kepler}. We find  a break in the RV occurrence around $\sim 2$--$3$\,au (Section \ref{sub:turnover}), close to the location of the snow line in our Solar System (e.g., \citealt{hayashi1981,demeo2014}), and show that a broken power law better describes the observed GP frequency as a function of orbital period. In Section \ref{sub:directimaging}, we demonstrate that such broken power law also explains the low occurrence of directly imaged giant planets. We compare the overall occurrence rate distribution with that predicted by different planet formation models and find good agreement with a subset of these models (Section \ref{sec:models}). Finally, we summarize our main results and discuss them in the context of the giant planets in our Solar System and the prominent structures detected in disks around young stars (Section \ref{sub:discussion}).

\section{Giant Planet Occurrence Rate} \label{sec:occurrence}
The intrinsic occurrence rate of planets can be calculated from the fraction of stars with detected planets in a survey and by making a correction for the number of non-detections. We calculate the average number of planets per star, hereafter occurrence rate, by averaging the inverse of the detection efficiency for each planet:
\begin{equation}
%\eta_\text{bin}
\eta = \frac{1}{n_\star} \Sigma^{n_\text{p}}_{\text{j}} \frac{1}{comp_\text{j}}
\end{equation}
where $comp_\text{j}$ is the survey completeness evaluated at the location of each planet $j$, the number of detected planets is $n_p$ and $n_\star$ is the number of surveyed stars.
The uncertainty on the occurrence rate is calculated from the square root of the number of detected planets per bin. Bin size is determined by dividing the period range (in log space) by the selected number of bins.

\subsection{Radial Velocity Occurrence Rate} \label{sub:rv}
We calculate the RV GP occurrence rate using the detected planets and completeness reported in \citet{mayor2011}. The RV sample in \citet{mayor2011} is a combination of the HARPS and CORALIE radial velocity surveys and includes a total of 822 stars and 155 planets. 

\begin{figure}[!h]
\centering
\includegraphics[scale=0.7]{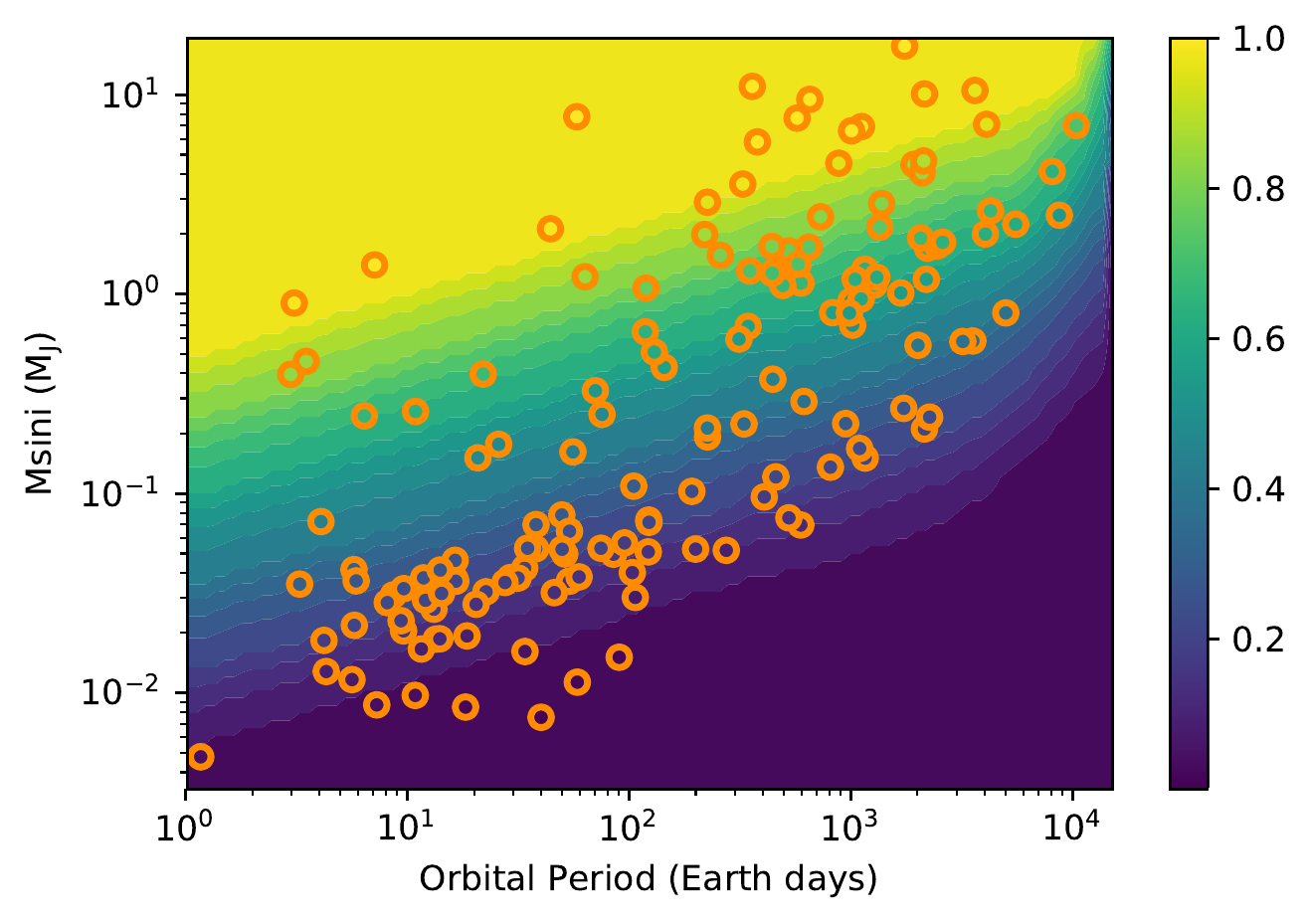}
\caption{RV survey completeness with color scheme given in the side bar. The completeness is calculated by linearly interpolating the completeness curves, which are from Figure~6 of \citet{mayor2011}. The orange circles represent the 155 planets from the HARPS+CORALIE survey. The planet list and detection efficiency are available online in electronic format as part of the \texttt{epos} package \citep{epos}. \label{fig:completeness}}
\end{figure}

We extract the survey completeness from Figure~6 in \citet{mayor2011}. This gives the probability that a planet with a given period P and minimum mass \textit{M\,{\rm sin}\,i} is detected and was calculated for each star and then averaged over all stars in the survey. Note that \citet{mayor2011} adopt circular orbits to estimate the exoplanet detectability as most of their planets have eccentricities below 0.5 and \citet{endl2002} have shown that eccentricities below this value do not substantially affect the RV detectability. Next, we recomputed the completeness over a finer grid by linearly interpolating on a uniform grid with \textit{M}sin\textit{i} between 0.001-20\,$M_{\rm J}$ and period between 1-20,000 Earth days (see Figure~\ref{fig:completeness}).

In our analysis, we consider planets with a minimum mass in the range 0.1-20\,$M_{\rm J}$, the lower value chosen to include all planets more massive than Neptune. We find that the GP occurrence rate increases with orbital period out to $\sim$1,000\,days (see Figure~\ref{fig:compare} dark green curve) which is consistent with previous results, e.g., \citet{mayor2011} and references therein. A more detailed analysis of the trend is described in Section~\ref{sub:turnover}. Note that the exact choice of upper and lower mass bins does not influence the broader trend described here (Figure~\ref{fig:compare} dark and light green curves). For the  number of planets per \textit{M}sin\textit{i} and orbital period bin, with corresponding completeness, see Appendix \ref{sec:occrates}. All data used in this paper and an example script to calculate planet occurrence rates are also available in the \texttt{epos} package \citep{epos}.

\subsection{Comparison with \textit{Kepler} Occurrence Rate} \label{sub:comparekep}
We calculate the GP occurrence rate from Kepler in a similar manner as RV. We use the planet candidate list and survey completeness from the latest DR25 \textit{Kepler} data release \citep{Mathur2017,thompson2018} as described in \citet{mulders2018}. 
Exoplanet eccentricities are not implemented in our study (or in the \texttt{epos} simulations), effectively assuming circular orbits. For the completeness, planets with small eccentricities (e$<$0.3) have also been assumed to be on circular orbits as any difference in the transit probability and duration is negligible, see   \citet{mulders2015b}. The completeness and number of planets per radius and orbital period bin is reported in Appendix \ref{sec:occrates}.

\begin{figure}%[!htb]
\centering
\includegraphics[scale=0.6]{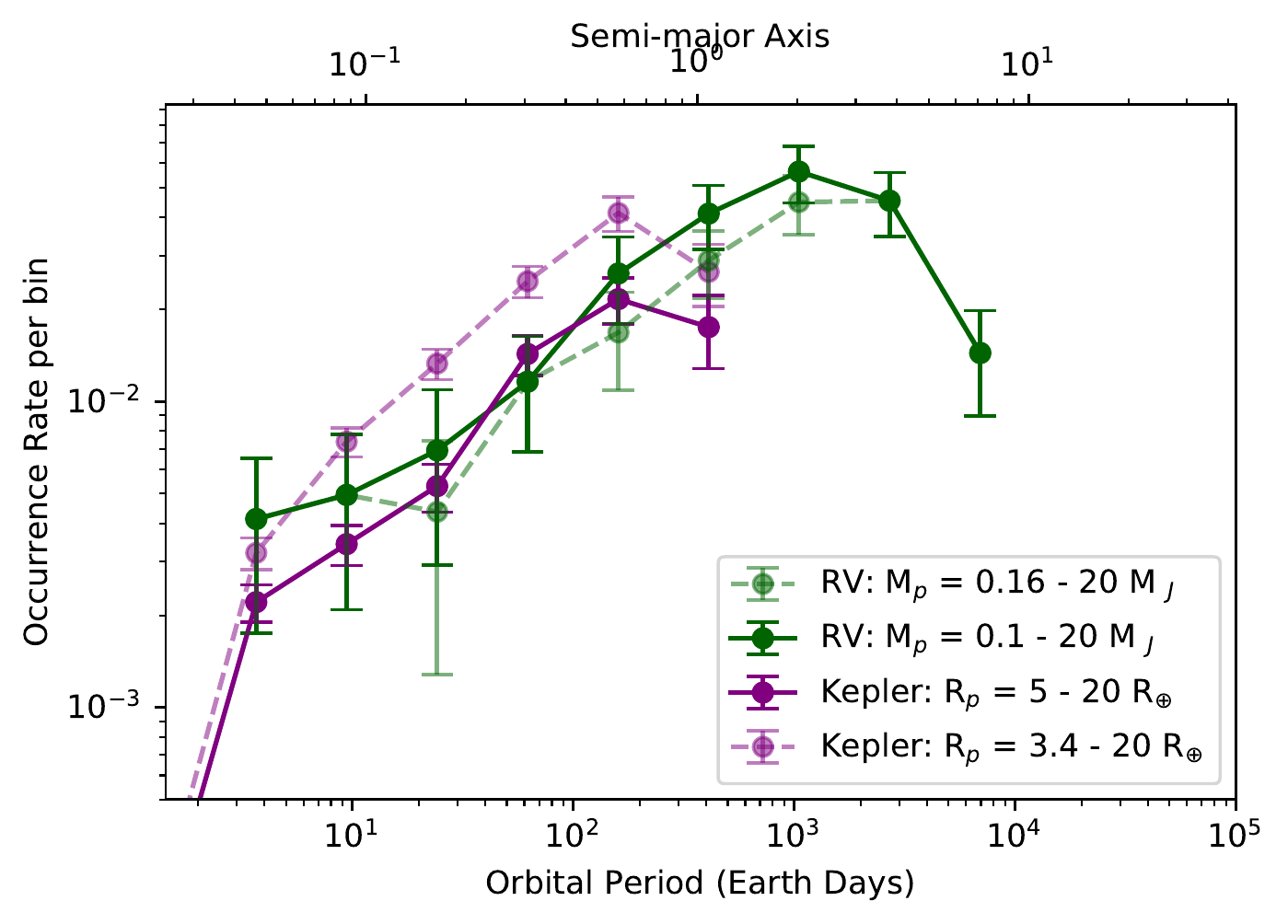}
\caption{GP occurrence rate as a function of orbital period (in days) for RV (dark solid green curve) and Kepler (dark solid purple curve) with the mass/radius ranges used in this paper. The pale dotted green curve represents the RV occurrence rate in the mass range used by \citet{mayor2011} whereas the pale dotted purple curve is for the \textit{Kepler} radius range used in the SAG13 study.  Note that the \textit{Kepler} pipeline is less complete, hence less reliable, in the longest period bin (300 - 1000 days), see e.g. \cite{schmitt2017} and \cite{thompson2018}.
\label{fig:compare}} 
\end{figure}

Since \textit{Kepler} measures planet radii and not masses, we use a mass-radius relation to select a planet radius bin that covers a similar RV planet mass bin. With the mass-radius relation in \citet{chen2016}, we set the lower value for the radius bin to 5\,$R_\earth$, as it is within 1$\sigma$ of the best-fit relation for 30\,$M_\earth$, while the upper value is set to 20\,$R_\earth$. Using a sample of planets with known masses and radii, \citet{lozovsky2018} have recently shown that planets with radii $>$4\,$R_\earth$ must have a significant H-He atmosphere (more than 10\%\ of the planetary mass). Hence, our choice of 5\,$R_\earth$ as the lower radius ensures that we include gaseous planets in our analysis. Note however that the trend of increasing occurrence with orbital period does not depend too much on the exact choice of radius bins (Figure~\ref{fig:compare}, light and dark purple curves).  

As can be seen in Figure~\ref{fig:compare}, the \textit{Kepler} and RV occurrence rates versus orbital period are very similar.  While Kepler and RV potentially probe different stellar populations as well as different planet size/mass regimes as described above, the rates are the same within 1 $\sigma$ for the bulk of the population which is beyond 10\,days. In the Hot Jupiter regime, i.e. inside 10\,days, we find that the \textit{Kepler} occurrence rate is lower (0.51$\pm$0.08\%) than the RV occurrence (0.9$\pm$0.5\%), as reported in previous studies (e.g., Figure~9 in \citealt{santerne2016}), though these values are consistent within 1 $\sigma$ (see also \citealt{petigura2018}). 

\begin{figure}%[!htb]
\centering
\includegraphics[scale=0.6]{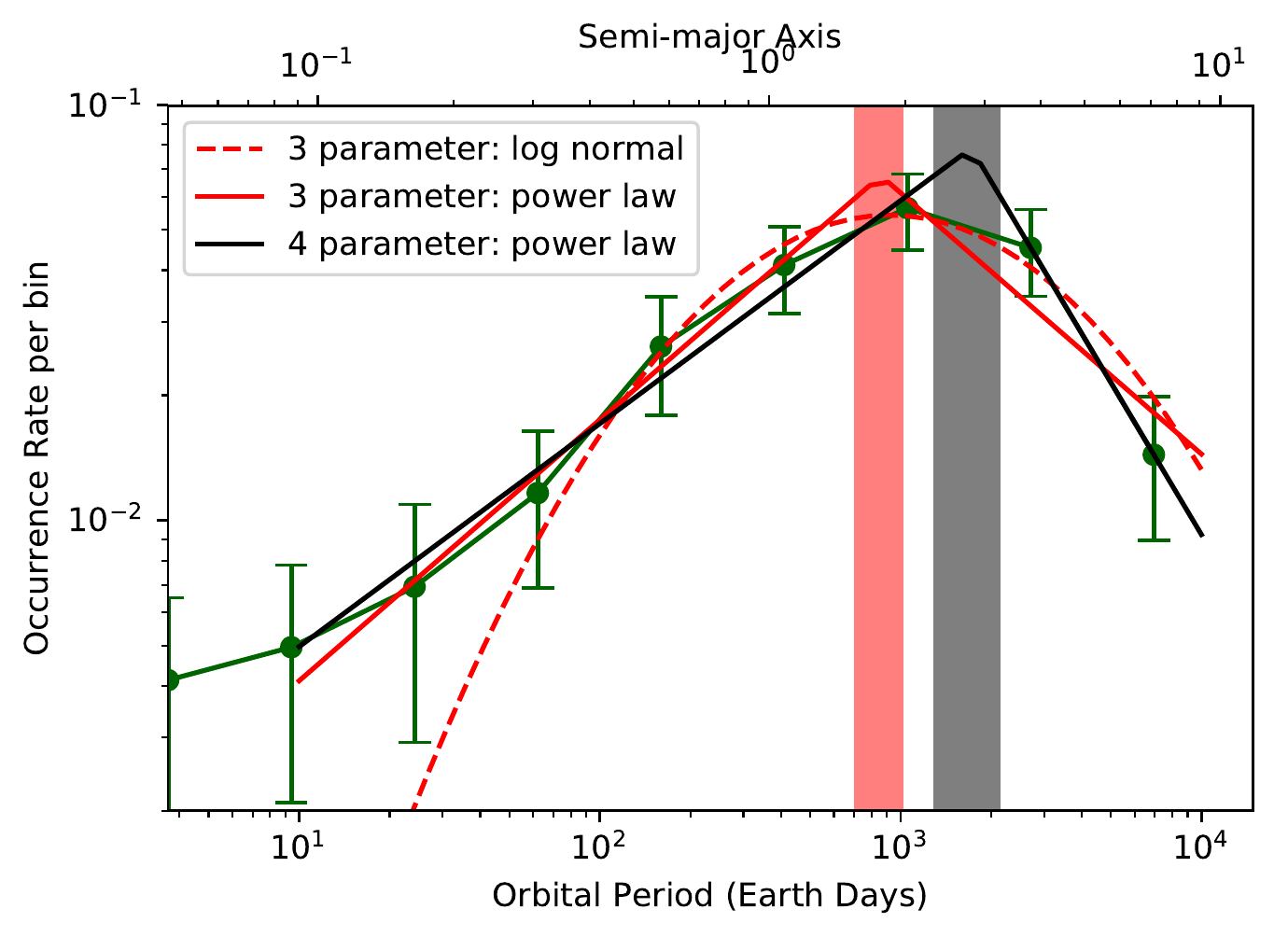}
\caption{Occurrence rate of 0.1-20\,$M_{\rm J}$ planets (green) with best fit relations beyond 10\,days: asymmetric broken power-law (solid black line), symmetric broken power-law (solid red line), and log-normal (dotted red curve). The location of $P_\text{break}$ is shown as a shaded region (gray for the asymmetric broken power-law and light red for the other two 3-parameter fits).
\label{fig:allfits}}
\end{figure}

\subsection{Turnover at $\sim$2.5 au} \label{sub:turnover}
As discussed in the previous subsections the occurrence rate of GPs increases with orbital period for periods where the RV and Kepler surveys overlap. However, beyond 1,000\,days the RV curve appears to have a turnover\footnote{Note that a turnover is also seen in the cumulative rate of giant planets (\textit{M}sin\textit{i}$ > 50 $M$_\oplus$) presented in \citet{mayor2011}, their Figure~8.}. We perform three statistical tests to characterize the break in the distribution.

\begin{table*}[!htb]
\centering
\begin{tabular}{ll|rrrrr} 
\hline\hline
Fit type & Function & \multicolumn{5}{c}{Parameters}  \\ 
\hline
 & $$ & $p_1$ & $P_\text{break}$ & $p_2$ & $m_1$ & Normalization\\ 
 & $$ & P $< P_\text{break}$ & in days & P $> P_\text{break}$ & & Constant \\ 
\hline
\multirow{3}{*}{\texttt{scipy}} & Asymmetric & $0.53\pm{0.09}$ & $1717\pm{432}$ & $-1.22\pm{0.47}$ & $-$ & $0.078\pm{0.014}$\\
& Symmetric & $0.63\pm{0.11}$ & $859\pm{161}$ & $-0.63\pm{0.11}$ & $-$ & $0.067\pm{0.012}$\\
& Log-normal & $-$ & $919\pm{105}$ & $-$ & $-$ & $0.084\pm{0.007}$ \\
\hline
\multirow{2}{*}{\texttt{epos}} & Asymmetric & $0.70^{+0.32}_{-0.16}$ & $2075^{+1154}_{-1202}$ & $-1.20^{+0.92}_{-1.26}$ & $-0.46\pm{0.06}$ & $0.83^{+0.19}_{-0.16}$\\
& Symmetric & $0.65^{+0.20}_{-0.15}$ & $1581^{+894}_{-392}$ & $-0.65^{+0.20}_{-0.15}$ & $-0.45\pm{0.05}$ & $0.84^{+0.18}_{-0.15}$\\
\hline\hline
\end{tabular}
\caption{Best fit parameters of the asymmetric, symmetric and log-normal distributions using \texttt{scipy} and \texttt{epos}.
\label{tab:parameters}}
\end{table*}

\begin{figure}%[!htb]
\minipage{0.5\textwidth}
  \includegraphics[width=\linewidth]{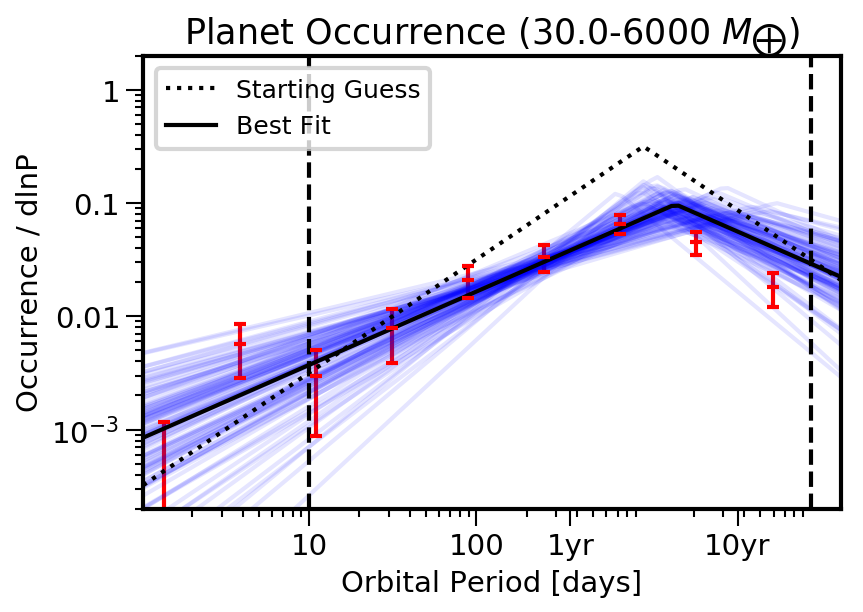}
\endminipage\hfill
\minipage{0.5\textwidth}
  \includegraphics[width=\linewidth]{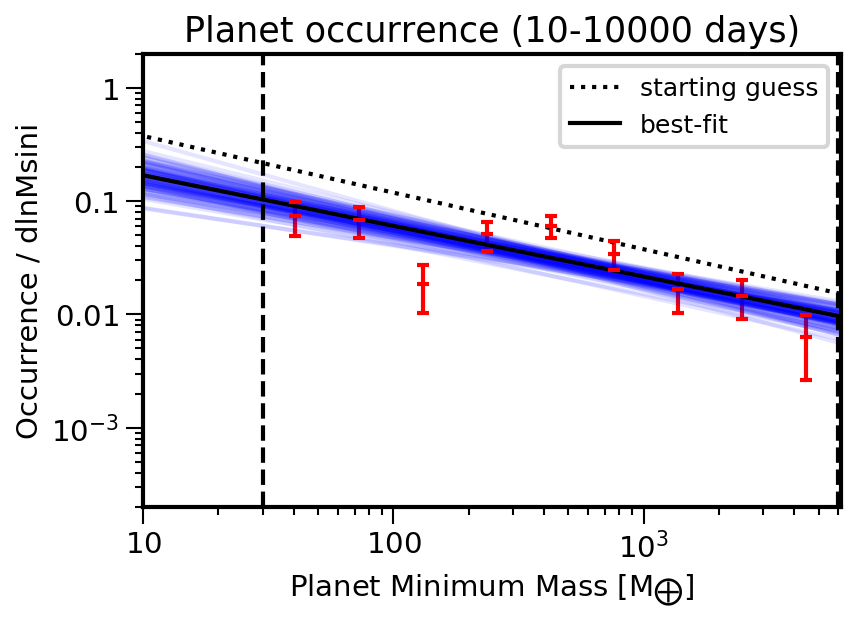}
\endminipage
\caption{\texttt{epos} posterior orbital period distribution (top) and planet mass distribution (bottom) for a symmetric power-law distribution in period. The red bars show the occurrence rates estimated using the inverse detection efficiencies for comparison. \label{fig:posterior} }
\end{figure}

First, we use the CRAN package \texttt{segmented} to evaluate the statistical significance of a breakpoint. \texttt{segmented} determines if an observed distribution can be best described by one or multiple linear segments. It does not use a grid search but rather an iterative procedure, starting only from possible breakpoints, and taking advantage of the fact that the problem can be linearized \citep{segmented1,segmented2}. It also uses a bootstrap restarting \citep{wood2001} to make the algorithm less sensitive to the starting values.
Since RV occurrence rates are typically better described by power-laws than linear functions (e.g., \citealt{cumming2008}), we fit the log of the RV occurrence rate per bin vs the log of the period. For planet masses 0.1-20\,$M_{\rm J}$ and periods $\sim 1-10,000$\,days, \texttt{segmented} finds a break at 1,766\,days ($\sim 2.8$\,au) and a probability that the distribution can be described by a single power-law (or line in log space) as low as 0.17\%\ using the associated davies test\footnote{The same tests applied to the 0.16-20\,$M_{\rm J}$ range used in \citet{mayor2011} find a break at $\sim$3\,au with a somewhat larger probability of 1.8\%\ that a single power-law can describe the RV occurrence between 1 and 10,000\,days.}. 

Motivated by these results, we then fit the occurrence rate and associated uncertainty with a broken power law in orbital period utilizing the optimize function from the \texttt{SciPy} package in Python:
\begin{equation}
\dfrac{d N}{d \log P} = A \begin{cases} 
\bigg(\frac{P}{P_\text{break}}\bigg)^{p_\text{1}} & if P \leq P_\text{break}\\
\bigg(\frac{P}{P_\text{break}}\bigg)^{p_\text{2}} & if P > P_\text{break}
\end{cases}
\label{eq:scipy}
\end{equation}
\\
where \textit{N} is the number of planets per period bin, \textit{P} is the orbital period, \textit{A} is a normalization factor, \textit{$P_\text{break}$} is the location of the break in the period distribution and, \textit{$p_1$} and \textit{$p_2$} are the power-law indices before and after the break, respectively. 

For an asymmetric broken power-law (4 parameters fit), $P_\text{break}$ is found to be between 1285$-$2149\,days (2.3$-$3.2\,au) with $p_1$=0.53$\pm$0.09 and $p_2$=-1.22$\pm$0.47\footnote{The results for a larger minimum mass of 0.16\,M$_{\rm J}$ are very similar:  $P_\text{break}$=1602$-$2086\,days while $p_1$=0.48$\pm$0.09 and $p_2$=-1.21$\pm$0.47 beyond.}. This best fit gives a low reduced $\chi^{2}$ of 0.12 suggesting that we might be over-fitting the data. Hence, we also consider a 3-parameter fit with a symmetric broken power-law in order to better constrain the slope after the break. In this case, $P_\text{break}$ is found to be closer in, between 698$-$1020\,days (1.5$-$2.0\,au), and $p_1$=$-p_2$=0.63$\pm$0.11, basically set by the larger number of data points inside $P_\text{break}$. This also gives us a low reduced $\chi^{2}$ of 0.11. 

A log-normal distribution, as that used for GPs around M dwarfs \citep{meyer2018}, returns a $P_\text{break}$ between 814$-$1024\,days. Given the symmetry of this function around the break point it is not surprising this $P_\text{break}$ is consistent with that obtained from the symmetric broken power-law fit. In this case the reduced $\chi^{2}$ is 0.74, higher than in the case of the power-law functions. 

Additionally, we use the Bayesian Information Criterion (BIC) test to determine which fit is statistically better. BIC is a criterion for model selection among a finite set of models in which the model with the lowest BIC is preferred. It is based, in part, on the likelihood function and is defined as BIC = n*ln(sse/n) + k*ln(n) where n is the number of observations, k is the number of variables and sse is the squared sum of the residuals. The bigger the difference in the BIC scores of two models (usually $>$2), the worse the model with the higher BIC score is. We find that the asymmetric (BIC score: -29.54) as well as the symmetric broken power law fit (BIC score: -31.62) were indeed better fits than the single power law fit (BIC score: 12.16) since they have significantly lower BIC scores. Figure~\ref{fig:allfits} provides a visual comparison of these best fit relations while Table \ref{tab:parameters} summarizes the best fit parameters.

Finally, we also use \texttt{epos}\footnote{\url{https://github.com/GijsMulders/epos}} version \texttt{v1.1} \citep{epos}, which uses the forward modeling approach described in \cite{mulders2018}, to constrain the occurrence rate of GPs. \texttt{epos} simulates exoplanet survey yields from an intrinsic distribution of planet properties by taking into account detection biases such as viewing inclination,
and constrains this distribution by adopting a Markov Chain Monte Carlo approach with the \texttt{emcee} Python algorithm \citep{foreman2013}.

For this analysis, we adjust \texttt{epos} for use with radial velocity surveys by constraining the distribution function of planet mass and orbital period directly from the observed survey data, without random draws.
We opt not to use the Monte Carlo simulation in \texttt{epos} because the number of detected planets in the RV survey is significantly lower than detected in the \textit{Kepler}, increasing the associated noise. Instead, we replace the steps where we generate a synthetic planet population by random draws and remove non-detectable planets to obtain a detected planet sample, by the following steps.

First, we adjust the planet distribution function from Equation \ref{eq:scipy}, to include a planet mass dependence.
\begin{equation}
f_M(M, P)= \dfrac{d^2 N}{d \log P d \log M} = c_\text{0} f(P) \bigg(\frac{M}{10 M_\oplus}\bigg)^{m_\text{1}}\
\label{eq:epos}
%M_\text{break} = 10 M_oplus
\end{equation}
where $c_0$ is a normalization factor, $f(P)$ is the broken-power law period distribution described in Equation~\ref{eq:scipy} (without the normalization factor $A$), and $m_1$ is the power-law index of the planet mass distribution $M$. The normalization factor is set such that the integral of the function over the simulated planet period and mass range equals the average number of planets per star.

%\begin{equation}
%f_\text{obs}(M, P)= f_\text{pl}(M, P) f_\text{det}(M, P)
%\end{equation}

Then, we convolve the true mass distribution with a function to take into account random viewing angles
\begin{equation}
g(M \sin i)= \frac{M \sin i}{\sqrt{M_T^2 - (M \sin i)^2}} % for M sin i < M
\end{equation}
where $M_T$ is the true planet mass
see Appendix \ref{sec:epos} for derivation, to obtain the simulated $M \sin i$ distribution $(f_M * g)$ with is function of $(M \sin i, P)$.

Last, we multiply this distribution with the detection efficiency, $f_\text{det}(M\sin i, P)$, to obtain a detectable planet distribution function:
\begin{equation}
f_\text{obs}(M\sin i, P)=
%(f_M(M, P)*g)(M \sin i) \cdot f_\text{det}(M\sin i, P)
(f_M*g)(M \sin i, P) \cdot f_\text{det}(M\sin i, P)
\end{equation}
We then compare this distribution function to the observed planet distribution, $\{M, P\}$, using a one-sample KS test to replace the two-sample KS test in \texttt{epos}.\footnote{We have verified with \texttt{epos} that using minimum mass instead of true mass leads to only a small underestimate in planet occurrence rates for such wide mass bins, only $\sim$12.9\% of the occurrence itself. We did this by fitting the \textit{M}sin\textit{i} distribution instead of the true mass distribution with \texttt{epos} and calculating the percentage change in the normalization factors.} We then proceed as in \citet{mulders2018} to identify the best parametric fit for the RV exoplanet population using \texttt{emcee}. A triangle plot of the best fit parameters for the symmetric power-law fit can be seen in Appendix \ref{sec:epos}.
This procedure has been implemented in version 1.1 of \texttt{epos}.

When modeling the period distribution with an asymmetric power-law, \texttt{epos} finds a break at $2075^{+1154}_{-1202}$ days with $p_1$=$0.70^{+0.32}_{-0.16}$  and $p_2=-1.20^{+0.92}_{-1.26}$ after the break, in agreement withing 1 $\sigma$ of our asymmetric power-law fit using \texttt{scipy}. The posterior distribution of planet orbital period and mass are shown in Figure \ref{fig:posterior}, for the symmetric \texttt{epos} fit. The corner plot showing the projections of the likelihood function can be found in Appendix \ref{sec:epos}. In the case of a symmetric power-law distribution, \texttt{epos} finds a break at $1580^{+894}_{-392}$ days with a slope of $p_1=-p_2=$ $0.65^{+0.20}_{-0.15}$ which is within 1 $\sigma$ of the values found with \texttt{SciPy} for a  symmetric power-law in period. The best fit power law index for the mass distribution is $\sim -0.45$ for both the symmetric and asymmetric power-law in period.\footnote{We increased the lower mass limit to 0.3\,M$_{\rm J}$ and 0.5\,M$_{\rm J}$ and found that there was no sigificant change in $p_1$ and $p_2$ with their respective $P_\text{break}$ at 1346$-$3470 days and 1504$-$3190 days which is consistent within 1 $\sigma$ of the 0.1\,M$_{\rm J}$ value.} Using the BIC test in \texttt{epos}, we found that the asymmetric (BIC score: 23.5) as well as the symmetric broken power law fit (BIC score: 18.9) were indeed better fits than the single power law fit (BIC score: 34.2) since it has a lower BIC score. Here, again, the broken power-law is  preferred over the single power-law since the difference in BIC scores is $>$10. A summary of our best fit values can be found in Table \ref{tab:parameters}. 

Hence, we conclude that there is evidence for a break in the RV GP occurrence rate, although the slope beyond the break is not well constrained because the RV data only extend to 10$^{4}$ days. An RV data set with a longer baseline is needed to put stronger constraints on the slope after the break. While we prefer a 3-parameter solution given the lack of observational constraints, where the slope is constrained mostly from the distribution before the break, as also motivated by \citet{meyer2018}, the fitted slope of the 4-parameter solution has a large uncertainty that is consistent with the 3-parameter solution to within 1 $\sigma$. The break has important implications when extrapolating GP occurrence rate at semi-major axis relevant for direct imaging surveys (Section \ref{sub:directimaging}). Additionally, it is important for our theoretical understanding of how and where giant planets form (see Section \ref{sec:models}).

\begin{figure}[!htb]
\centering
\includegraphics[scale=0.6]{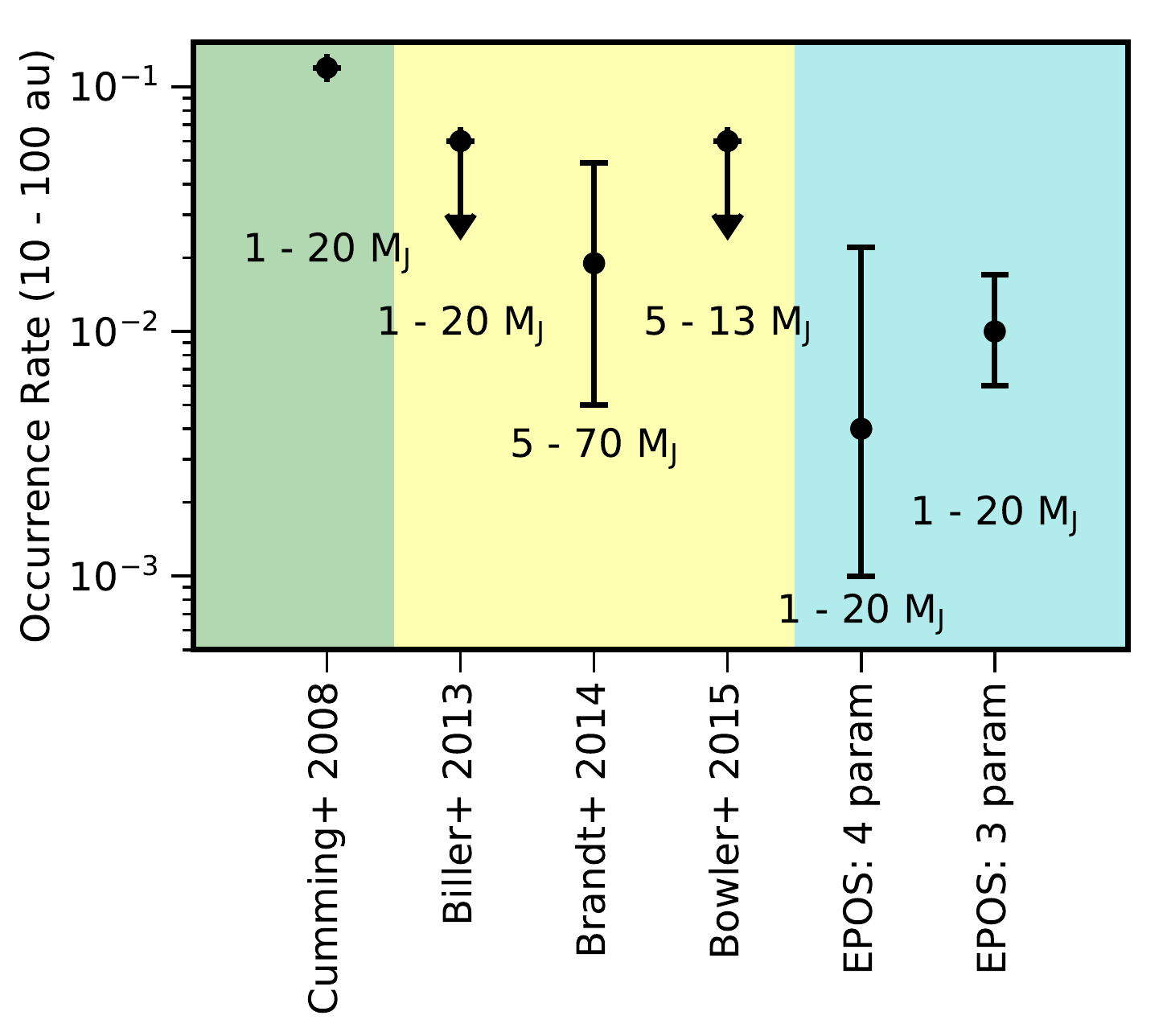}
\caption{Extrapolated and observed direct imaging rates at 10-100\,au. The green panel represents the occurrence rates extrapolated using the single power-law fit to RV occurrence curves such as \citet{cumming2008}. The yellow panel represents the observed occurrence rates from direct imaging surveys such as \citet{biller2013}, \citet{brandt2014} and \citet{bowler2015}. The blue panel represents the extrapolated occurrence rates using 3 and 4 parameter broken power-law fits to RV occurrence rates using \texttt{epos} \citep{mulders2018}. \textbf{Note 1.} Upon extrapolation of the SAG13 baseline function to the same distances, we get an occurrence rate of 85.4\%\ which is higher than any of the predicted rates as well as the ones calculated in this paper. \textbf{Note 2.} The direct imaging rates (yellow panel) have a hidden mass distribution, which if assumed to be flat, cannot directly be compared to our rates.
 } \label{fig:extrapolate}
\end{figure}

\begin{table*}%[!htb]
\centering
\begin{tabular}{ll|r|rr|rr} 
\hline\hline
Fit type & Function & Free parameters & \multicolumn{4}{c}{Direct Imaging Prediction (\%)}  \\ 
\hline
 & $$ & $$ & (a) Giants & (b) Giants & (c) Jupiters & (d) Jupiters \\ 
 & $$ & $$ & $0.1-13 ~M_{\rm J}$  & $0.1-13 ~M_{\rm J}$ & $1-13 ~M_{\rm J}$ & $1-13 ~M_{\rm J}$ \\ 
 & $$ & $$ & $10-100$ au  & $<20$ au & $10-100$ au & $<20$ au \\ 
\hline
\multirow{4}{*}{\texttt{scipy}} & asymmetric & 4  & 4.30 & 8.9 & 2.1 & 8.8 \\
& symmetric & 3 & 5.94  & 7.9 & 2.9 & 7.8 \\
& single & 2 & 24.2 & 14.2 & 11.7 & 6.8  \\
& log-normal & 3 & 5.49 & 8.4 & 1.3 & 3.7 \\
\hline
\multirow{4}{*}{\texttt{epos}} & asymmetric & 5 & $1.7^{+7.9}_{-1.3}$ & $16^{+4}_{-3}$ & $0.4^{+1.8}_{-0.3}$ & $4.6^{+1.0}_{-0.7}$  \\
& symmetric & 4 & $3.9^{+2.9}_{-1.7}$ & $17^{+3}_{-3}$ & $1.0^{+0.7}_{-0.4}$ & $4.9^{+0.7}_{-0.6}$ \\
& single & 3 & $74.2^{+24.7}_{-17.0}$  & $32^{+7}_{-5}$ & $17.4^{+4.1}_{-3.0}$ & $8.5^{+1.2}_{-1.1}$\\
\hline
Cumming+08 & single & 4 & 14 & - & 6.8 & 17-20 \\
Kopparapu+18 & single & 4 & - & 101 & - & -  \\
\hline\hline
\end{tabular}
\caption{Comparison of extrapolated GP occurrence rates for (a) $0.1-13 ~M_{\rm J}$ between 10 - 100\,au. (b) $0.1-13 ~M_{\rm J}$ within 20\,au  which were calculated using the SAG13 mass range (converted from radius) and period range. (c) $1-13 ~M_{\rm J}$ between 10 - 100\,au which were calculated assuming $d N / d \log \textit{M\,{\rm sin}\,i} = \text{constant}$ (in the case of \texttt{scipy} only). (d) $1-13 ~M_{\rm J}$ within 20\,au which were calculated using the semi-major axis range used in \citet{cumming2008} but the fit is extrapolated out to 20\,au in order to compare with the extrapolated SAG13 rates from \citet{kopparapu2018}
\label{tab:fits}}
\end{table*}

\subsection{Predictions for Direct Imaging} \label{sub:directimaging}
Direct imaging surveys are mostly sensitive to planets at large ($>$10 au) separations. Such surveys have found that GPs more massive than $\sim$5\,$M_{\rm J}$ are rare, with an occurrence of only 1\%, at separations between a few tens to a few hundreds au (e.g., \citealt{bowler2016,galicher2016}). This value is lower than the RV GP occurrence rate inside a few au, see e.g. Figure~1. Integrating within a period of 2,000\,days \cite{cumming2008} reported an occurrence of $\sim$10\%\ for 0.3-10\,M$_{\rm J}$, an order of magnitude higher than the rate of directly imaged planets. Hence, it was realized early on that the single RV power law in semi-major axis cannot extend at large separations (e.g., \citealt{kasper2007,chauvin2010,nielsen2010}). 
Recently, \citet{bryan2016} conducted an RV+imaging survey of stars with already known exoplanets and found that the frequency of  GP companions declines with semi-major axis beyond 3-10\,au. Here, we show that the turnover we find in the RV occurrence rate at $\sim$2.5\,au naturally explains the high occurrence of GPs within a few au and the low occurrence rate of planets further out.

Our analysis of the \citet{mayor2011} RV planets, including their survey completeness, recovers previous results obtained with a single power law (see Appendix \ref{sec: rates}) and, most importantly, suggests that the GP occurrence does not increase with orbital period beyond $\sim$3\,au (see Section \ref{sub:turnover}). Using functions that take into account this turnover, we calculate yields at large semi-major axis and find that they are significantly lower than those predicted by a single power law, see Table~\ref{tab:fits}. 

For instance, when we extrapolate the \texttt{scipy} asymmetric and symmetric power laws at the location where direct imaging is most sensitive to, i.e. 10-100\,au, and assuming a flat planet mass dependence between 0.1-13\,$M_{\rm J}$, we predict a GP occurrence rate of 4.3\%\ and 5.9\%\, respectively (column (a) in Table~\ref{tab:fits}). \texttt{epos} finds similar values, albeit the uncertainties on the occurrence rate are large for the 4-parameter fit due to the uncertainty in the slope after the break. 

In the same period range, the occurrence of Jupiters, planets with masses between 1 and 13\,$M_{\rm J}$, is even lower (column (c) in Table~\ref{tab:fits}), well in agreement with that from direct imaging surveys. On the other hand, for GPs across all the mass and period ranges in Table~\ref{tab:fits}, the occurrence rates estimated using a single power law in period are an order of magnitude to several orders of magnitudes higher than those obtained with broken power laws as well as a log-normal fit.

A visual summary of GP occurrence rates in the semi-major axis range directly relevant to direct imaging (10$-$100\,au) is shown in Figure~\ref{fig:extrapolate}. The figure includes the extrapolated rate using the single power law in period reported in \citet{cumming2008}, three representative occurrence rates from direct imaging surveys \citep{biller2013,brandt2014,bowler2015}, as well as the \texttt{epos} values reported in Table~\ref{tab:fits}. As can be seen in the figure, the \texttt{epos} rates obtained with a broken power law in period are within 1$\sigma$ of the direct imaging values from \citet{brandt2014} and consistent with the upper limits reported in \citet{biller2013} and \citet{bowler2015}. The extrapolated rate from \citet{cumming2008} using a single power-law is too high, clearly inconsistent with the low occurrence of GP reported by direct imaging surveys. Hence, we can conclude that the broken power-law fit is more consistent with the observed GP occurrence rates at larger orbital periods. 

Recently, \cite{kopparapu2018} extrapolated the \textit{Kepler} SAG13 occurrence rates between $1.6- 20$\,au to evaluate the GP yield of future direct imaging missions. As the extrapolation is based on the assumption that a single power law in period describes well the GP occurrence (see also Appendix \ref{sec: rates}), the yield is very large, basically each star has a GP between $\sim$2-20\,au (see column (b) in Table~\ref{tab:fits}). However, if a broken power-law (or log-normal) distribution better describes the GP occurrence rate, Table~\ref{tab:fits} shows that the occurrence of $0.1-13$\,$M_{\rm J}$ planets is about a factor of $\sim$5 lower than that reported in \cite{kopparapu2018}. This has important implications for the science goals that can be achieved by future direct imaging missions.

\section{Comparison with theoretical models} \label{sec:models}
As shown in Section \ref{sec:occurrence}, the occurrence rate of GPs rises with orbital period, peaks between 2-3\,au, and decreases beyond. Such a feature could be an imprint of giant planet formation and/or subsequent evolution in the disk. Within the core accretion paradigm, giant planet formation happens as a two-step process: first a solid core with a critical mass of order 10\,M$_\earth$ must form, then the rapid accretion of a massive gaseous envelope sets in. As GPs form in a gaseous disk, they must subject to gas-driven migration through tidal interaction with their nascent disk. As in most models they form beyond the snow line, inward migration likely plays a role in shaping the semi-major axis distribution of GPs at short orbital periods. However, it is also possible that the GPs could form in situ (e.g., see \citealt{batygin2016}) or, under favorable conditions, they could migrate outward (e.g., see \citealt{dangelo2012}).

As the time scale for inward type-II migration is typically shorter than the disk lifetime, GPs will be accreted onto the star if migration is not stopped.
There are several mechanisms proposed to halt the inward migration and create the observed population of warm giants: interaction with other GPs; photoevaporation carving a hole in the disk; or slow type-II migration that have been included in population synthesis models.

In this section, we make a qualitative comparison between planet occurrence rates and planet population synthesis models that employ these different physical mechanisms. 
We use the grid of surviving planets from \citet{jennings2018comparative} and \citet{ida2018}, and two grids from the \texttt{DACE} database based on the simulations in \citet{mordasini2018}. For each model, we calculate the predicted occurrence rate as the fraction of simulated star systems that form a planet with mass larger than 0.1\,$M_{\rm J}$ in each period bin in order to directly compare to the curves we computed in Section \ref{sec:occurrence}.

\begin{figure}%[!htb]
\minipage{0.5\textwidth}
  \includegraphics[width=\linewidth]{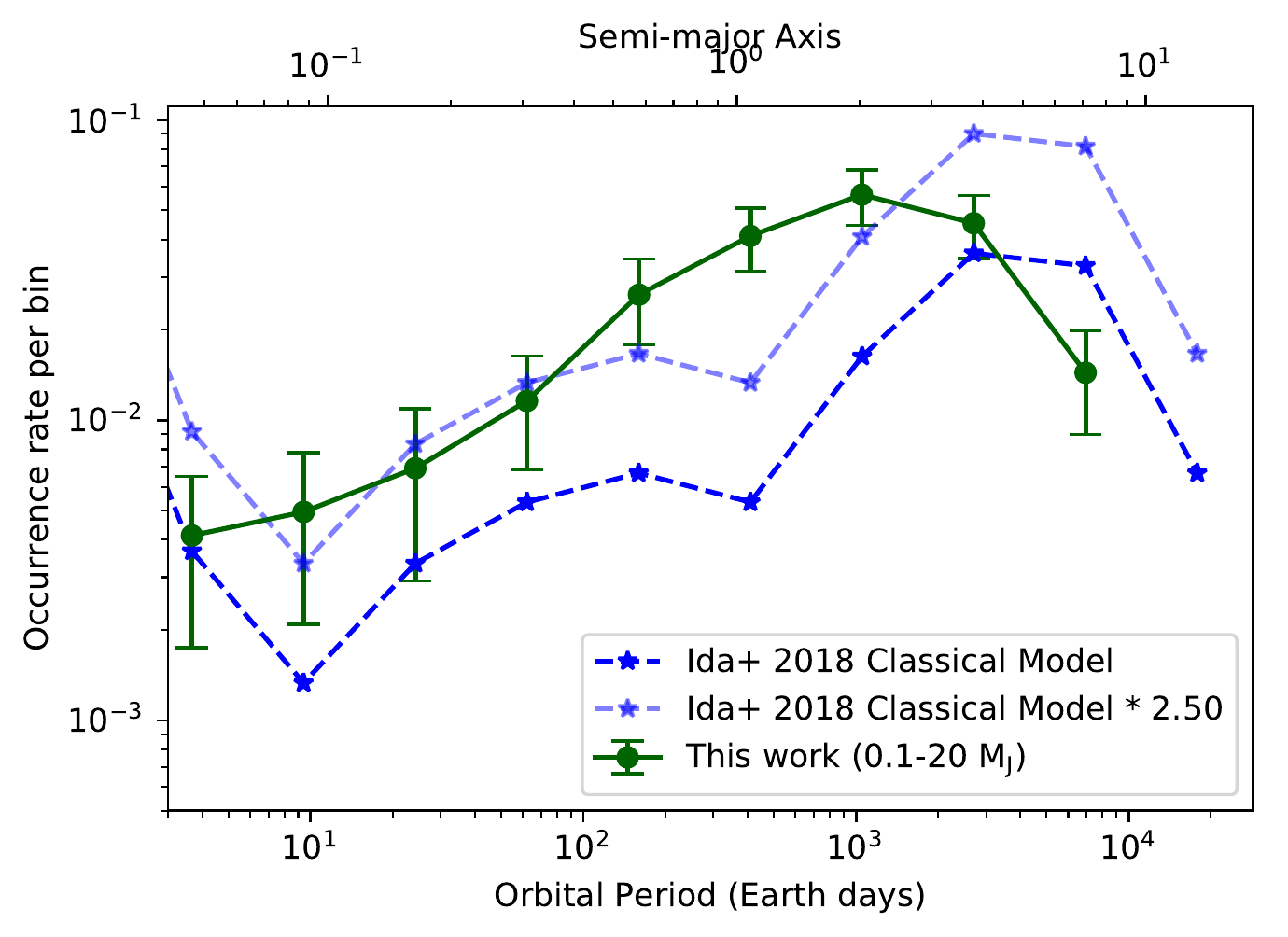}
\endminipage\hfill
\minipage{0.5\textwidth}
  \includegraphics[width=\linewidth]{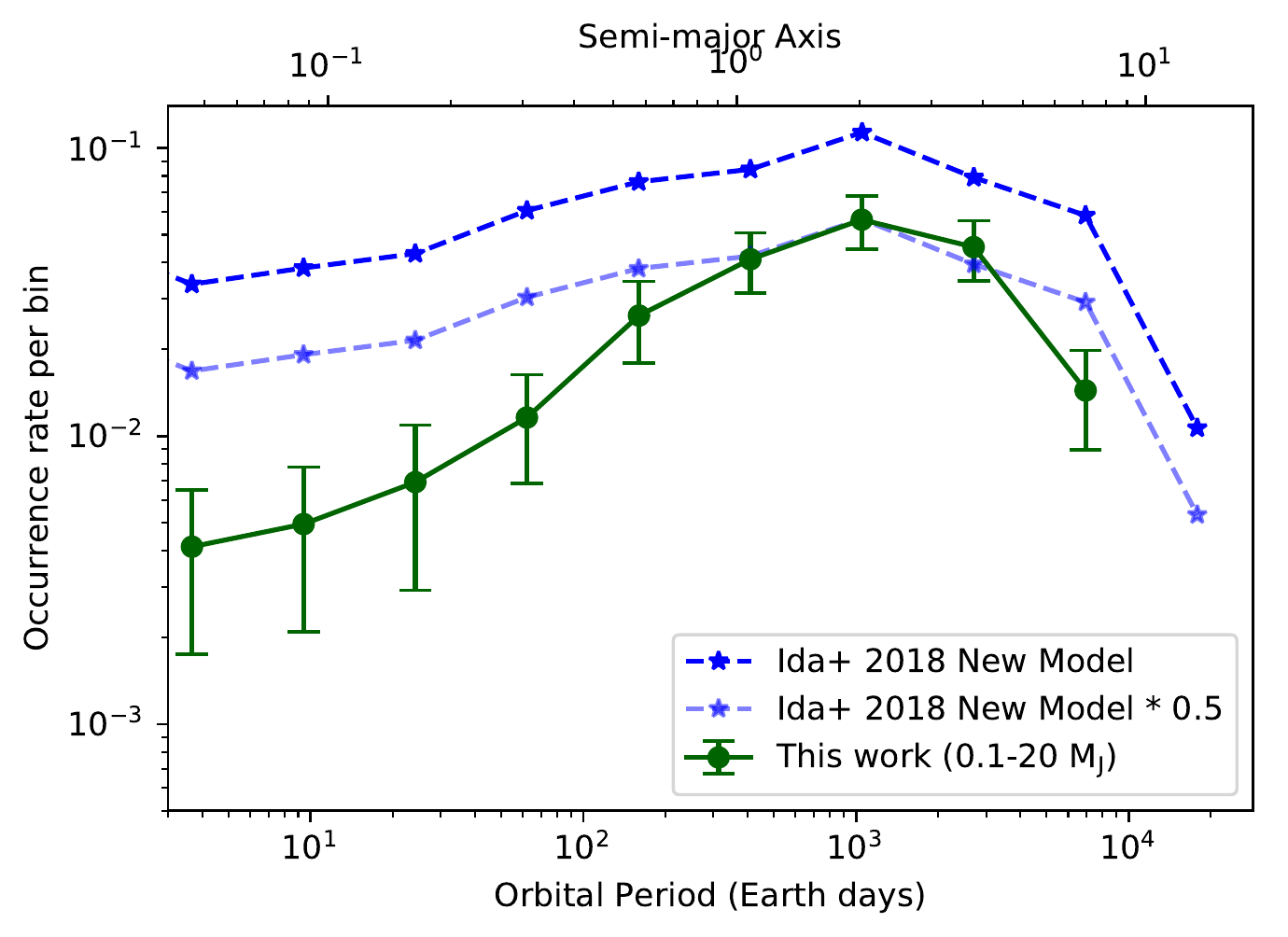}
\endminipage
\caption{Comparison of the \citet{ida2018} model (dark blue) and the RV occurrence rate (green curve). 
\textbf{Top:} Classical model scaled up by a factor of 2.5 (light blue) to show the similarities in the slope of both distributions. \textbf{Bottom:} The new model scaled down by a factor of 0.5 (light blue) to show the overlap in the region where both curves turnover. 
\label{fig:idalin}}
\end{figure}

\subsection{Formation via core accretion: Ida et al. 2018} \label{idalin}
\citet{ida2004a,ida2004b} were the first to carry out core accretion population synthesis models. They used the results of N-body simulations to model the accretion phase of cores starting from planetesimals and Kelvin-Helmholtz contraction for the accretion of gas onto cores. 
While Type~I  migration is not included, protoplanets large enough to open a gap are subject to Type~II migration, and move inward.

Recently, \citet{ida2018} published updated synthetic planet populations for two different implementations of Type II migration, classical and new. The classical model assumes that Type II migration is associated with gas accretion through the disk, as in \citet{ida2004a,ida2004b}. The new model is based on recent high-resolution simulations by \citet{kanagawa2018} showing that there is a disconnect between the migration of the gap-opening planet and the disk gas accretion, resulting in a reduced migration rate. The slower Type-II migration results in a larger fraction of planets being retained at short orbital periods. 

As can be seen in the top panel of Figure \ref{fig:idalin}, the classical model indeed underpredicts the number GPs between 10-1000 days by a factor $\sim$2.5. However, the slope of the predicted orbital-period distribution in this region is similar to the estimated planet occurrence rates (dark green curve). There is no clear indication of a turnover within the observed range ($<10^4$\,days). On comparing the scaled model to the data, we get a high $\chi^{2}$ value of 186.65, thus implying that this model is not a good fit to the data.

The orbital period distribution of the new model (see Figure \ref{fig:idalin}, bottom panel) is much flatter than observed. The new model overproduces the number of giant planets at all orbital periods, in particular those at the shortest orbital periods. We scaled down the model to better fit the peak of the RV distribution. The slope of the orbital period distribution is much flatter than the observed one. The new model does shows a turnover at a period of $\sim 1,000$ days that matches well with the observed break in the planet occurrence rate distribution. Here, we get a $\chi^{2}$ value of 91.84 hinting that this new model is a better fit than the classical one.

As the observed RV distribution lies in between the new and classical model, a direct comparison between model and observations may be used to calibrate the strength of Type II migration.

\begin{figure}%[!htb]
\minipage{0.5\textwidth}
  \includegraphics[width=\linewidth]{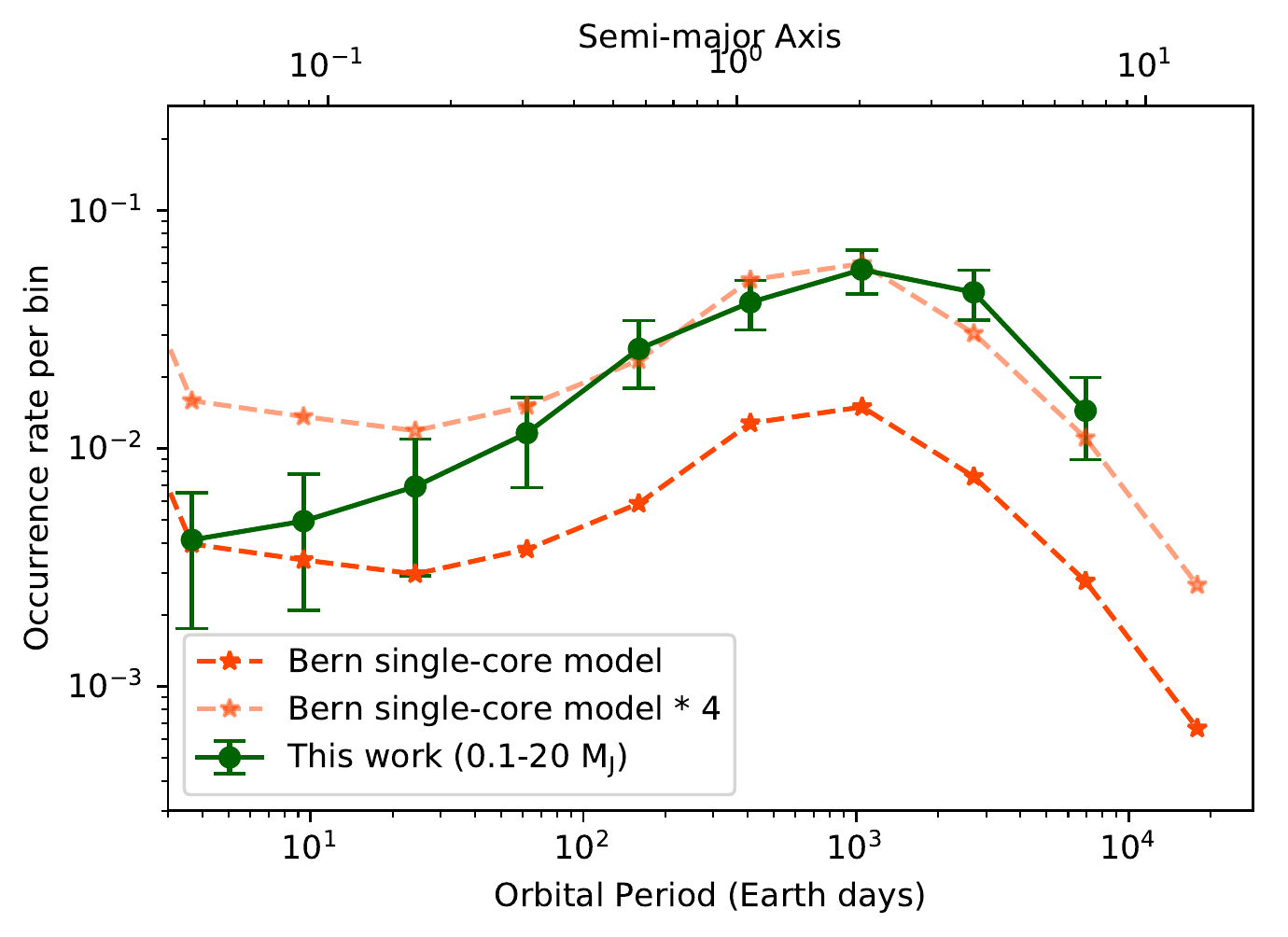}
\endminipage\hfill
\minipage{0.5\textwidth}
  \includegraphics[width=\linewidth]{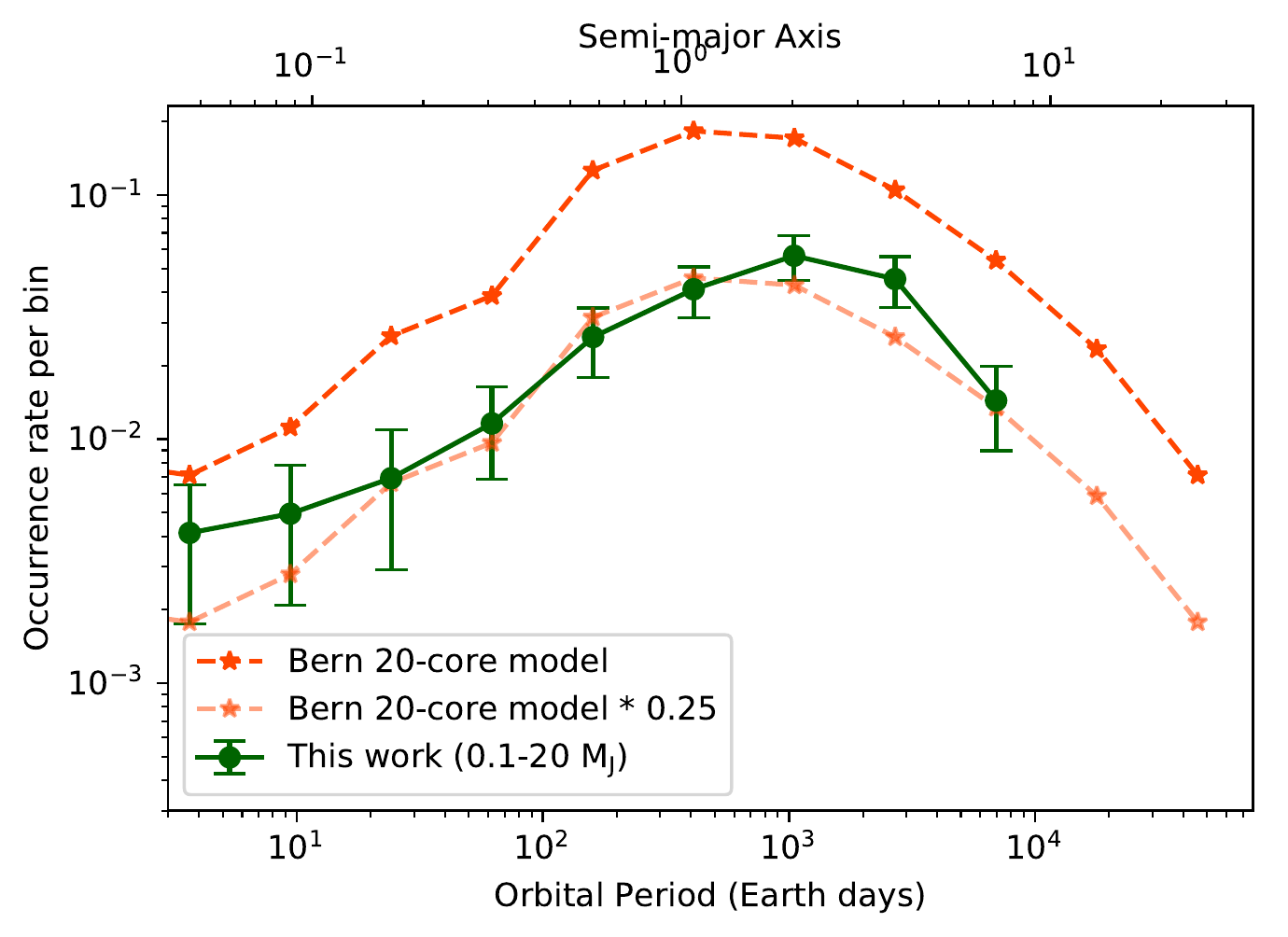}
\endminipage
\caption{Comparison of the Bern model (dark orange dashed curve) and RV occurrence rate (in green). In light orange we show: the single-core model scaled up by a factor of 4 (top panel); and the 20-core model scaled down by a factor of 0.25 (bottom panel).
\label{fig:mordasini}}
\end{figure}

\subsection{Formation via core accretion and multi cores interaction: The Bern Model} \label{sub:mordasini}

As in \citet{ida2004a}, the GP formation model used in \citet{mordasini2018} relies on the core accretion paradigm. However, these more recent population synthesis include updated prescriptions for the evolution of the protoplanetary disk as well as for migration, including Type~I migration of cores that are not massive enough to open gaps \citep{dittkrist2014} and directly calculate the N-body interaction of concurrently forming protoplanets \citep{alibert2013}.

In this contribution, we compare two populations obtained from an updated version of the \citet{mordasini2018} model. Each population comes from a different type of model: one model includes just one planetary embryo that can grow into a GP (single-planet) while the other includes multiple embryos that can form multiple GPs (multi-planet). The multi-planet model additionally uses a N-body module to study concurrent growth and interactions of multiple planetary embryos that are injected into each disk. The results from these simulations will be presented in more details in Emsenhuber et al. (in prep.).

The single-planet simulations predict an increase in the GP occurrence rate between 100$-$1000\,days (top panel in Figure \ref{fig:mordasini}), as observed (dark green curve in the same figure). However, they under-predict the overall number of GPs at those orbital periods by a factor of $\sim$4 when compared to the RV occurrence rate. When scaled to match the observed RV peak, the single-planet models over-predict the number of GPs inside $\sim$30\,days but the turnover beyond $\sim$1,000\,days is very similar to the observed one giving us a $\chi^{2}$ value of 39.12.

The GP occurrence rate for the multi-planet simulations with initially 20 embryos per disk lower panel of Figure~\ref{fig:mordasini}) has a similar shape as the observed one, i.e. increasing occurrence with semi-major axis and a turnover after which occurrence rate decreases. In order to match the occurrence rate, results need to be scaled down by a factor of 0.25, implying that only 1 in 4 stars would form the simulated systems. On comparing the scaled model to the RV data, we get a low $\chi^{2}$ value of 6.96, suggesting that the scaled model is a good fit to the data.

The two models have overall the same shape, except for the inner planets. The difference in the total occurrence rate follows from the increase of the number of embryos in the multi-planet model, which increases the chance of forming GPs. The overall similar shape of the single- and multi-core models tell us that the planet-planet interactions do not significantly affect the GP formation process. We expect however that a further increase of the number of embryos could affect these results.

The single-planet model overproduces inner planets whereas the multi-planet underproduces them. In the single-planet model, inward migration of GPs is very efficient, which causes the pile-up of HJs. For the multi-planet model, migration is less efficient, possibly due planet-planet interactions which helps stabilizing GPs which are in mean motion resonances. 

\begin{figure}%[!htb]
\centering
\includegraphics[scale=0.61]{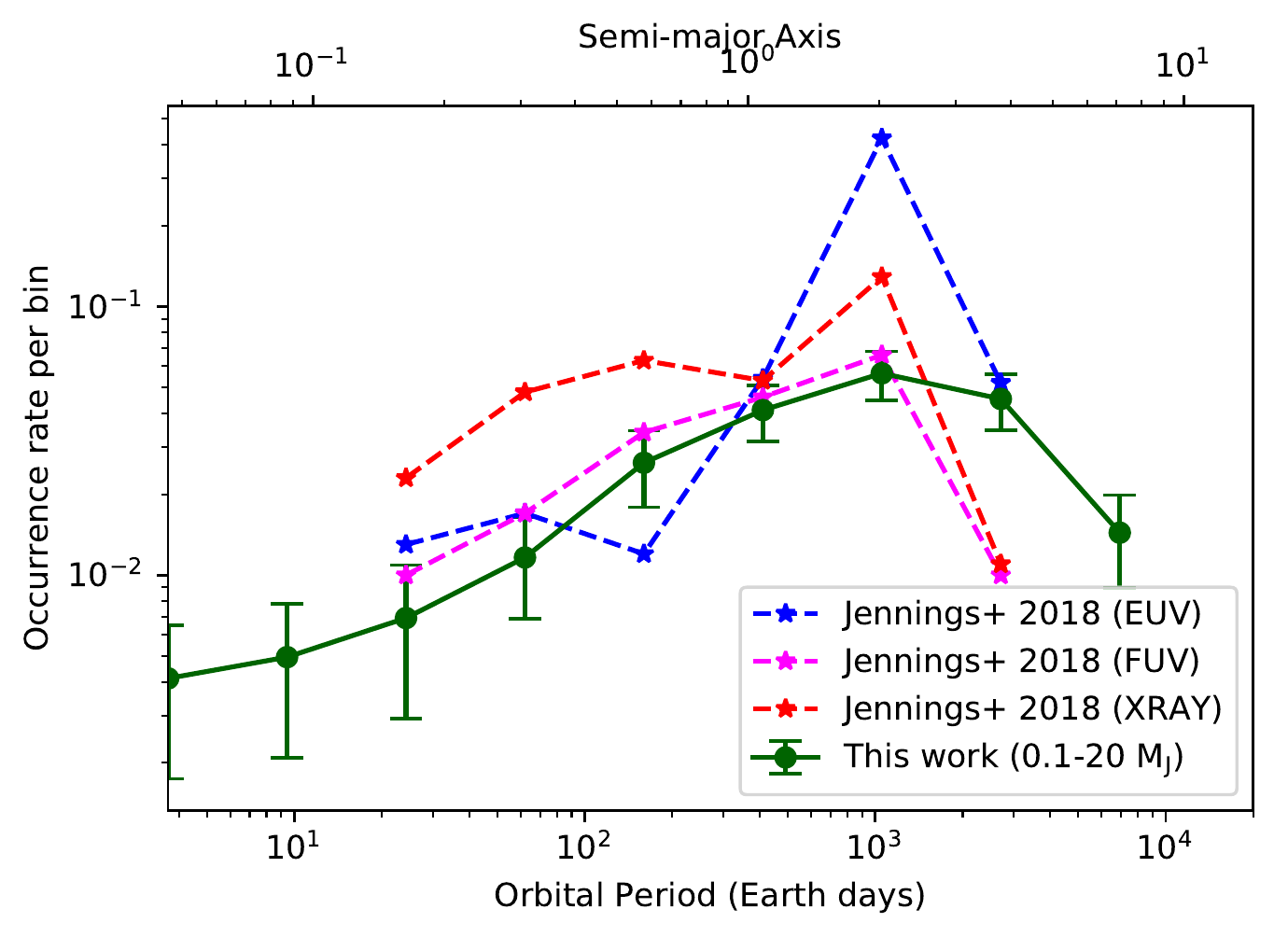}
\caption{Comparison of \citet{jennings2018comparative} models with the RV occurrence rates calculated in this paper (solid green curve). In these models, disk dispersal is driven by stellar EUV (blue), FUV (magenta), or X-ray (red) photons through so called photoevaporative winds.
\label{fig:jennings}}
\end{figure}

\subsection{GP trapping due to photoevaporation disk clearing: Jennings et al. 2018} \label{sub:jennings}
The models by \citet{jennings2018comparative} explore the late evolution of fully formed GPs in viscously evolving and photoevaporating disks, similarly to \citet{alexander2012deserts} and \citet{ercolano2015}. Unlike \citet{ida2004b,ida2004a} and \citet{mordasini2018}, they do not simulate the formation of planetary cores nor the subsequent atmospheric accretion. 
However, they do have a more detailed treatment of the disk clearing phase.

The simulation of \citet{jennings2018comparative} begins with a viscously evolving disk irradiated by either a pure EUV, X-ray, or FUV-dominated stellar flux. A planet ranging in mass from 0.5 to 5\,$M_{\rm J}$ is inserted at 5\,au at a random time between 0.25\,Myr and the time of disk clearing. The planet moves inward through Type~II migration, further accretes mass as gas flows across its gap, while the disk viscously evolves and is being photoevaporated. After a few Myrs, the mass accretion rate falls below the wind photoevaporative rate, hence a gap opens in the disk which can slow down or completely stop the migration of giant planets. The location of the gap, and hence the final location of planets, depends on the stellar high energy photons driving photoevaporation, i.e. $\sim$0.8\,au for EUV, $\sim$1.7\,au for X-ray and $\sim$4\,au for FUV. 
 
\citet{jennings2018comparative} end their simulations either when the disk surface density is $\leqslant$ $10^{-8} g$ $cm^{-2}$ or when the migrating planet reaches a separation  $\leqslant$0.15\,au from the star. A thousand of these simulations are conducted for each of the three photoevaporative profiles, by varying only the photoevaporative mass loss rate, planet formation time, and initial planet mass.

On comparison with the RV occurrence rates (Figure~\ref{fig:jennings}), we find that the FUV model (magenta curve) has an occurrence rate that best matches the observed rates and slope between 30-1000\,days with a $\chi^{2}$ value of 29.12. The X-ray model (red curve) does a fairly good job for the GP occurrence at $\sim$1,000\,days but tends to over-predict the number of GPs at shorter orbital periods ($\chi^{2}$ value of 288.56). Finally, the EUV model  over predicts the increase in planet occurrence around 1,000\,days or shows an increase in planet occurrence between 100-1000\,days that is steeper than observed, after scaling to the peak. Since EUV creates a gap closer in than the FUV or X-ray case, the viscous timescale to drain the inner disk is shorter, i.e. a surviving planet will be more likely stalled at the gap opening location which creates a larger pile-up than observed. The EUV model gives an extremely high $\chi^{2}$ value of 1947.84 meaning that it is not a good match to the data.
All models show a break at $\sim$1,000\,days but a steeper drop than the data beyond. 
However, the steep drop likely results from no planets being injected outward of 5\,au.  

\section{Summary \& Discussion} \label{sub:discussion}
We computed the giant planet occurrence rates out to $10,000$ days based on the planet detections and simulated survey completeness from the HARPS and CORALIE radial velocity surveys \citep{mayor2011}.
We characterize the shape of the orbital period distribution and evaluate our findings in the context of directly imaged planets and planet formation models. 
We find that:
\begin{enumerate}
\item The occurrence of giant planets from \textit{Kepler} (radii $5-10$\,R$_\earth$) and RV (masses 0.1-20\,M$_{\rm J}$) show the same rising trend with orbital period for periods between $10$--$100$\,days. As pointed out in the past (e.g., \citealt{santerne2016}), the {\it Kepler} occurrence of hot Jupiters ($<$10\,days) is about half of the RV one but we show that the value is within 1$\sigma$ of the large uncertainty associated with the RV data.
\item There is evidence for a break in the GP occurrence rate around $\sim$1,000-2,000\,days. For solar-mass stars these periods correspond to semi-major axis $\sim 2$--$3$\,au, bracketing the location of the snow line in the Solar System.
\item The break in GP occurrence rate decreases the giant planet yields when extrapolated to orbital periods accessible by direct imaging surveys. Using \texttt{epos} we calculate an occurrence of $1^{+0.7}_{-0.4}\%$ for planets more massive than Jupiter between 10-100\,au, in agreement with the values reported from direct imaging surveys. 
\item Different planet population synthesis based on core accretion and including Type-II migration produce a turnover in the GP occurrence rate around the snow line. We find that models with multiple planet cores per disk qualitatively seem to be a better match to the observed distribution.

\end{enumerate}
In a recent study \citet{wittenmyer2016} used their Anglo-Australian Planet Search survey to estimate that only $6.2^{+2.8}_{-1.6}\%$ of solar-type stars have a Jupiter analog, i.e. a giant planet with masses between 0.3-13\,M$_{\rm J}$ located between 3 and 7\,au. In the same planet mass and semi-major axis range, we derive an occurrence of $3.8\pm0.8\%$ from the \texttt{epos} best fit symmetric power law, in agreement with \citet{wittenmyer2016} within the quoted uncertainties. Thus, it appears that Jupiter analogs are rather rare.

Using the same best fit model, we also calculate an integrated frequency for planets between 0.1$-$100\,au of $26.6^{+7.5}_{-5.4}\%$ for 0.1-20\,$M_{\rm J}$ and $6.2^{+1.5}_{-1.2}\%$ for planets more massive than Jupiter (1-20\,$M_{\rm J}$).
These statistics are interesting in the context of structures recently identified in protoplanetary disks. For example, \citet{van2016} analyzed an unbiased sample of disk candidates based on \textit{Spitzer} catalogs and retrieved a frequency of dust cavities larger than 1\,au in radius of
23\% . While about half of them can be explained as the result of late disk evolution and dispersal by star-driven photoevaporation \citep{ercolano2017}, the other half are more likely to host one or more GPs.  As this  statistic is lower than our integrated occurrence of 0.1-20\,$M_{\rm J}$,  there appears to be enough mature GPs to explain the frequency of transition disks.
Rings and narrow gaps are very common in disk surveys biased toward the brightest millimeter sources, with an occurrence as high as 85\% from the ALMA DSHARP survey  \citep{dsharp1,huang2018}. However, this number goes down to $\sim$38\% in the Taurus disk survey from \citet{long2018} which covers fainter millimeter disks, hence it is more representative of the entire disk population. Considering that the fraction of disks in Taurus is 75\% \citep{luhman2009}, the fraction of structures from this survey is then $\sim$28\%. 
This value is similar to our occurrence of 0.1-20\,$M_{\rm J}$ planets between 0.1-100\,au. Hence, there is no disagreement between the count of mature exoplanets and disk structures {\it if} only one GP is necessary to reproduce all the observed structures in each disk and migration re-distribute the forming GPs over  a large range of semi-major axis, from 0.1 out to 100\,au. However, if multiple structures cannot be explained by one GP alone, it would be important to explore if planets less massive than 0.1\,$M_{\rm J}$ could open such gaps as their occurrence is larger than that of GPs both inside as well as outside the snow line \citep{suzuki2016,pascucci2018}.

Going forward, these occurrence rate distributions can be used to update planet yield estimates for future missions and provide context for \textit{Gaia}, which is expected to detect over 20,000 high-mass ($\sim 1-15$\,M$_{\rm J}$) planets out to $\sim$5\,au  (e.g., \citealt{perryman2014}). The large sample size from \textit{Gaia} may reveal structures around and beyond the snowline and place more stringent constraints on planet migration and formation models.

\acknowledgements
R.B.F. would like to thank Kyle A. Pearson for his valuable insight and expertise while recalculating the \citet{mayor2011} completeness. We would also like to thank Shigeru Ida and Jeff Jennings for sharing simulated planet populations. This paper includes data collected by the \textit{Kepler} mission. Funding for the \textit{Kepler} mission is provided by the NASA Science Mission directorate. This material is based upon work supported by the National Aeronautics and Space Administration under Agreement No. NNX15AD94G for the program Earths in Other Solar Systems. The results reported herein benefited from collaborations and/or information exchange within NASA’s Nexus for Exoplanet System Science (NExSS) research coordination network sponsored by NASA’s Science Mission Directorate.

\software{\texttt{NumPy} \citep{numpy}, \texttt{SciPy} \citep{scipy}, \texttt{Matplotlib} \citep{pyplot}, \texttt{emcee} \citep{foreman2013}, \texttt{corner} \citep{corner}, \texttt{epos} \citep{epos}, \texttt{CRAN segmented} \citep{segmented1,segmented2}}

\clearpage
\appendix
\section{Occurrence Rates} \label{sec:occrates}
\begin{figure}[!htb]
\centering
\minipage{0.8\textwidth}
  \includegraphics[width=\linewidth]{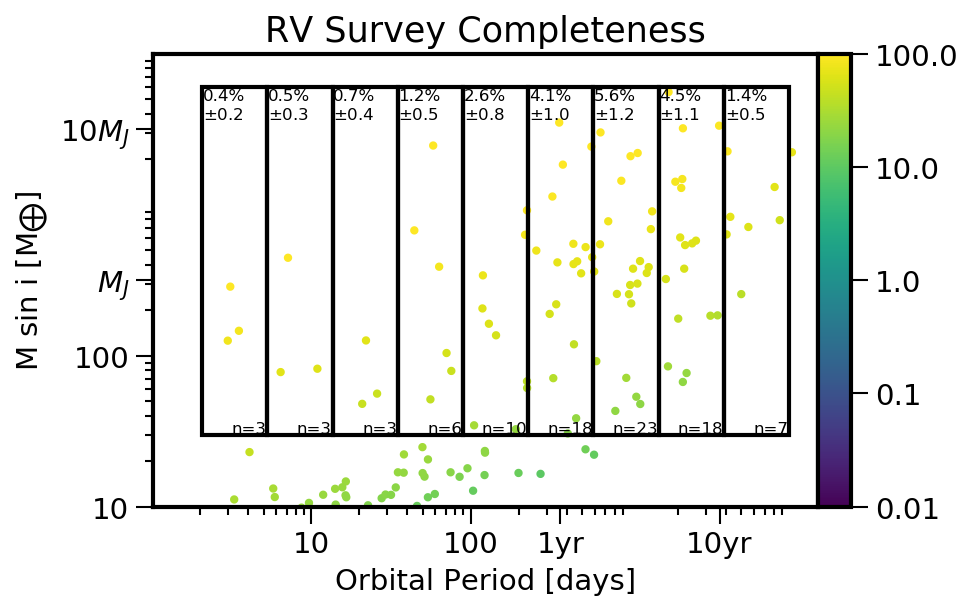}
\endminipage\hfill
\minipage{0.8\textwidth}
  \includegraphics[width=\linewidth]{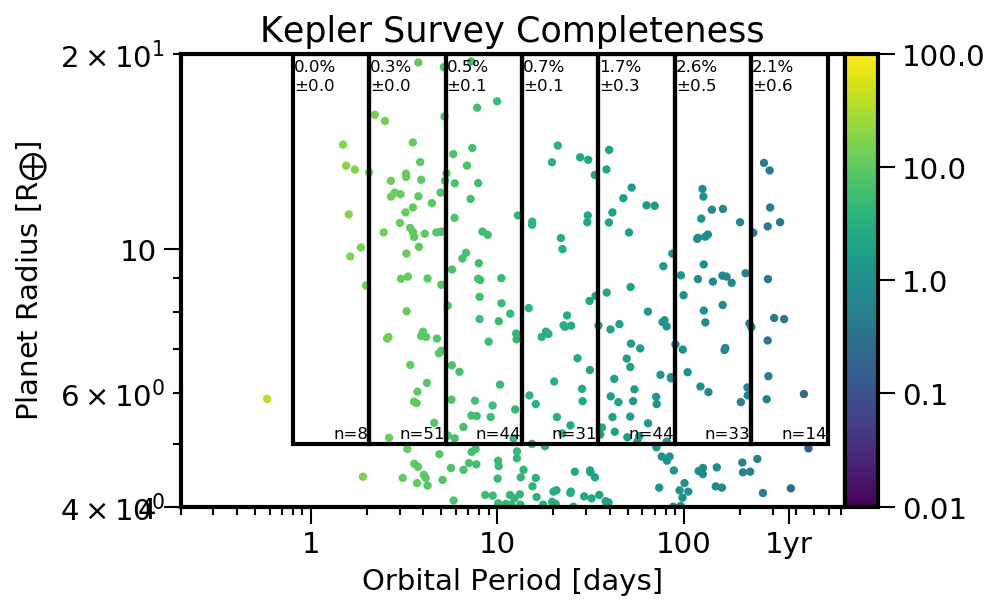}
\endminipage
\caption{\textbf{Left:} Planet Occurrence for RV survey per mass and orbital period bin. The color bar represents the completeness per planet.
\textbf{Right:} Planet Occurrence for \textit{Kepler} \texttt{dr25} survey per radius and orbital period bin. Dots show planet candidates, color-coded by the survey completeness (in percent) at that location. 
\label{fig:occrates}}
\end{figure}

Figure \ref{fig:occrates} shows the \textit{Kepler} and RV survey completeness and the number of planets per radius/\textit{M}sin\textit{i} and orbital period bin.

\clearpage

\section{\texttt{epos} parametric fit} \label{sec:epos}
\begin{figure}[!htb]
\centering
\includegraphics[scale=0.4]{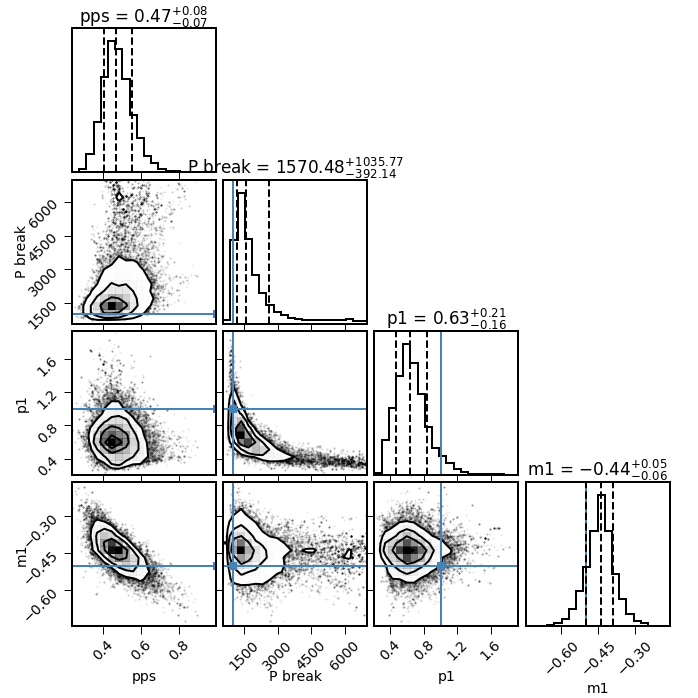}
\caption{\texttt{epos} corner plot showing the projections of the likelihood function for the five parameters that define the 2D broken power-law occurrence rate vs. orbital period and planet mass. Blue lines indicate the initial guess. The corner plot was generated using the open-source \texttt{Python} package \texttt{corner} by \citet{foreman2016}.
\label{fig:triangle}}
\end{figure}

Figure \ref{fig:triangle} shows the best fit parameters and associated uncertainties with a run for a symmetric power-law fit in period and a single power-law in \textit{M}sin\textit{i} that used 50 walkers for 1000 Monte Carlo iterations and a 200-step burn-in.

The conversion from planet mass distribution to $M \sin i$ distribution goes as follows.
For a planet of mass $M$ the measured minimum mass is $m= M \sin i$.
For random viewing angles, the inclination is distributed across $[0,\pi/2]$ as 
\begin{equation}
g(i)= \sin i. 
\end{equation}
The distribution of minimum mass is then:
\begin{equation}
g(m)= g(i) \frac{d i}{d m}
\end{equation}
and substituting $i=\sin^{-1}(m/M)$ gives
\begin{eqnarray}
g(m)= \sin(\sin^{-1}(m/M)) \frac{d}{dm}(\sin^{-1}(m/M))\\
= (m/M) \frac{1}{\sqrt{1-(m/M)^2}}\\
= \frac{m}{\sqrt{M^2-m^2}}
\end{eqnarray}
for $m \leq M$ and $g(m)=0$ if $m>M$.

\clearpage
\section{Over-prediction at large orbital periods when using a single power-law} \label{sec: rates}
\begin{figure}[!htb]
\centering
\plottwo{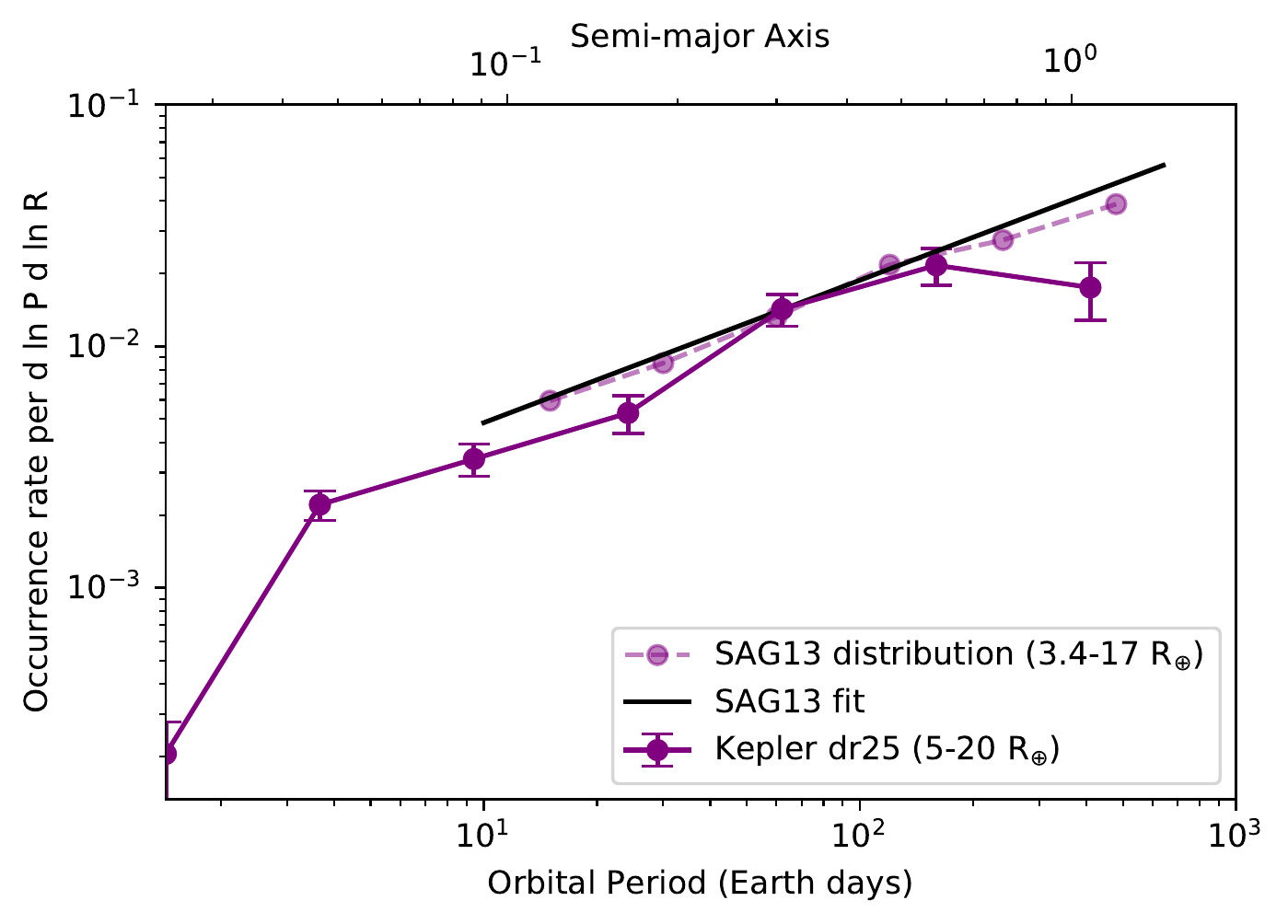}{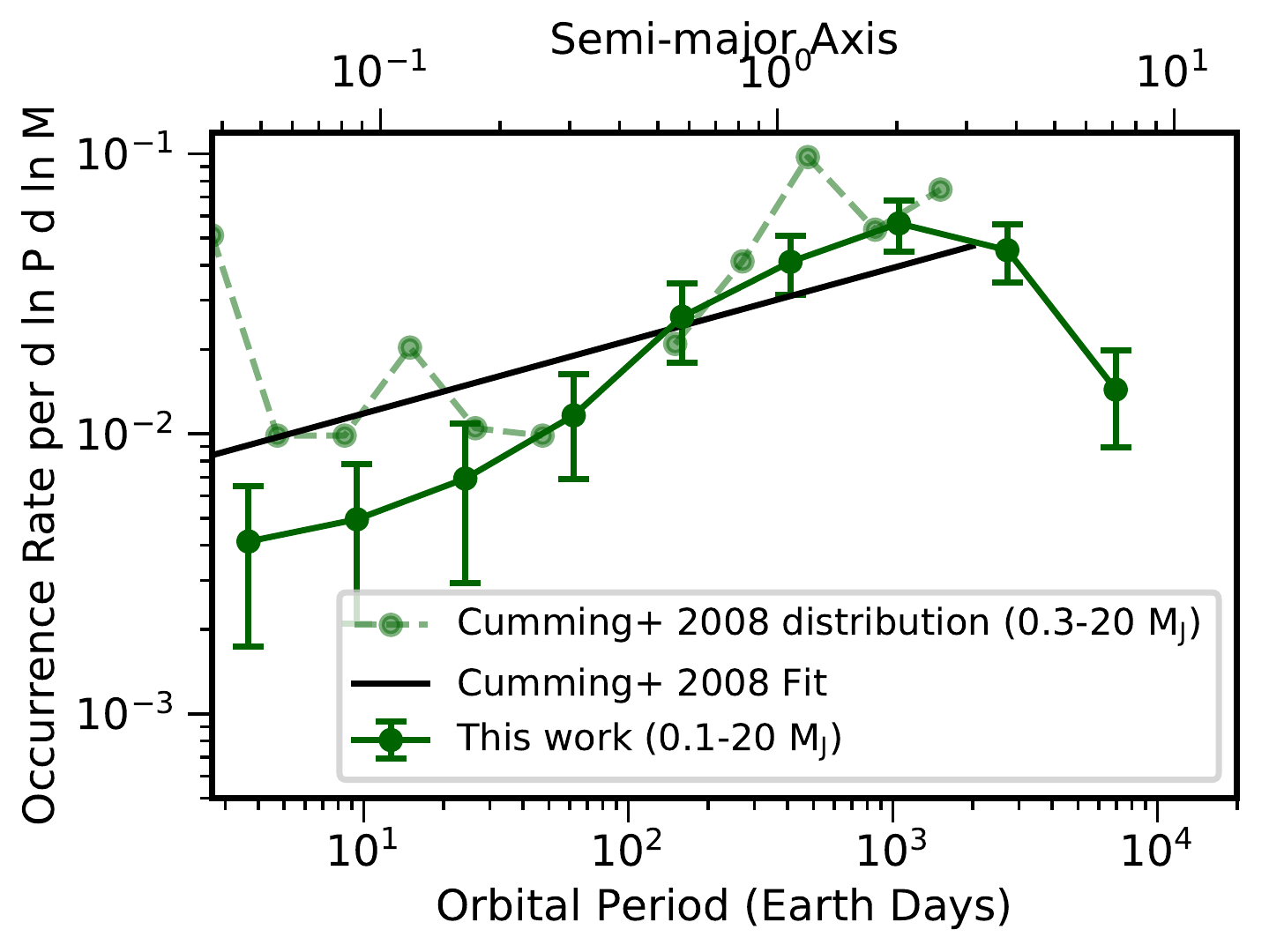}
\caption{\textbf{Left:} SAG13 fit shown in comparison with \textit{Kepler} dr25 occurrence curve (solid purple curve) calculated in this paper. The pale dotted purple curve is the SAG13 occurrence curve with different radius bins (3.4$-$17 $R_\earth$) to the ones used in this paper (5$-$20 $R_\earth$). \textbf{Right:} \citet{cumming2008} fit shown in comparison with the RV occurrence curve (solid green curve) calculated in this paper. The pale dotted green curve is the \citet{cumming2008} occurrence curve with different mass bins (90-6000 $M_\earth$) to the ones used in this paper (30-6000 $M_\earth$). 
\label{fig:fits}}
\end{figure}

The NASA's Exoplanet Science Analysis Group-13 (SAG-13) collected occurrence rates from different teams. The occurrence of GP (3.4 - 17 $R_\earth$) is fitted with a single power law within 10 - 640 Earth days. Similarly, \citet{cumming2008} have used a single power-law in order to explain the GP population of the RV data for GPs (0.3 - 20 $M_J$) out to 2,000 days. These single power-laws have been extrapolated in recent literature to predict the occurrence of GPs at large orbital distances. An example of these fits can be seen in Figure \ref{fig:fits}. The left panel shows the fit to SAG-13 distribution (3.4 - 17 $R_\earth$). In order to evaluate the planet yield of direct imaging missions, \citet{kopparapu2018} further extrapolate this fit between 1.6$-$20\,au to get a GP occurrence of 101\%.

\citet{cumming2008} extrapolate the single power-law distribution to estimate the number of planets within 20\,au to be 17$-$19\%. They find the slope of the period distribution to be 0.26 $\pm$ 0.1 as compared to a slope of 0.53 $\pm$ 0.09  (before the break) that we find when we fit a broken power law to the RV distribution. We were also able to reproduce the \citet{cumming2008} value for the slope when we fit a single power-law to the RV curve. Upon extrapolation of the \citet{cumming2008} power-law between 10 - 100\,au, we get an occurrence rate of 25\%. 

Both the \citet{cumming2008} and the SAG-13 extrapolated values are much higher that those from direct imaging observations (see Figure \ref{fig:extrapolate}) as well as those calculated using a broken power-law in this paper (see Table \ref{tab:fits}). 

\clearpage

\bibliographystyle{apj}
\bibliography{ref}

\end{document}